\documentclass[aps, twocolumn,superscriptaddress, longbibliography, nofootinbib]{revtex4-2}
\usepackage[hidelinks,colorlinks=true,linkcolor=blue,citecolor=blue]{hyperref}
\usepackage{lipsum,adjustbox}
\usepackage{ifthen}
\usepackage{forest}
\usepackage{nicefrac}
\usepackage{braket}
\usepackage{tikz}
\usepackage{dsfont}
\usepackage{amsthm}
\usepackage{ulem}
\usepackage{bm}
\usepackage{amsmath}
\usepackage{amssymb}
\usepackage{physics}
\usepackage{graphicx}
\usepackage{colortbl}
\usepackage{xcolor}
\usepackage{soul}
\usepackage{algorithm}
\begin{document}
\title{Sub-SQL electronic field sensing by simultaneously using quantum entanglements and squeezings}
\author{X. N. Feng}
\affiliation{Information Quantum Technology Laboratory, International Cooperation Research Center of China Communication and Sensor Networks for Modern Transportation, School of Information Science and Technology, Southwest Jiaotong University, Chengdu 610031, China}
\author{M. Zhang}
\affiliation{School of Physical Science and Technology, Southwest Jiaotong University, Chengdu 610031, China}
\author{L. F. Wei\footnote{E-mail: lfwei@swjtu.edu.cn}}
\affiliation{Information Quantum Technology Laboratory, International Cooperation Research Center of China Communication and Sensor Networks for Modern Transportation, School of Information Science and Technology, Southwest Jiaotong University, Chengdu 610031, China}

%%%%%%%%% ABSTRACT
\begin{abstract}
Quantum entanglement and quantum squeezing are two most typical approaches to beat the standard quantum limit (SQL) of the sensitive phase estimations in quantum metrology. Each of them has already been utilized individually to improve the sensitivity of electric field sensing with the trapped ion platform, but the upper bound of the demonstrated sensitivity gain is very limited, i.e., the experimental 3dB and theoretical 6dB, over the SQL. Here, by simultaneously using the internal (spin)-external (oscillator) state entanglements and the oscillator squeezings to effectively amplify the accumulation phase, we show that these sensitivity gains can be effectively surpassed. Hopefully, the proposal provides a novel approach to the stronger beaten of the SQL for the sensitive sensings of the desired electric field and also the other metrologies.
\end{abstract}
\maketitle
% \thispagestyle{empty}
%%%%%%%%% BODY TEXT %%%%%%%%%%%%%%%%%%%%%%%%%%%%%%%%%%%%%%%%
%%%%%%%%% BODY TEXT %%%%%%%%%%%%%%%%%%%%%%%%%%%%%%%%%%%%%%%%

%\section{Introduction}
{\it Introduction,---}
In recent years, how to use the quantum resources to beat the SQL of the parameter estimation sensitivity~\cite{Caves, Nagata2007,Pezze,Anisimov2010,Giovannetti2011,Jaewoo2011,Hosten2016,Liu2020,Zhuang2020} has been paid much attention. Specially, with the experimental trapped ion platform, various approaches have been investigated to implement the desired sensitive sensings of the weak electronic fields~\cite{Gilmore2018,Burd,Wolf2019,Gilmore,Affolter2023,Maiwald2009,Knunz2010,Biercuk2010,Ivanov2015, Liu2021,Liu2022, Gilmore2018,Burd,Wolf2019,Gilmore,Affolter2023,Backes2021}, as the response of the external vibration of the trapped ion is very sensitive to the applied weak electric field. For example, a sensitive sensing of the weak electric force has been experimentally demonstrated with very high sensitivity at $1yN/\sqrt{Hz}$-level~\cite{Maiwald2009}, which however is still limited by the SQL.
Therefore, utilizing the quantum resources, typically such as entanglement and squeezing, to implement the sensitive sensings of the weak electric field beyond the SQL is particularly desirable.

Recently, it has been demonstrated that, the sub-SQL weak force sensitivity can be achieved by using either the squeezing on the external vibrational state~\cite{Burd} or the spin-oscillator entanglement~\cite{Gilmore}. However, the sensitivity gain over the SQL (which is practically related to the involved mean phonon number) can only reach $3dB$ for the experimental demonstrations, although the theoretical prediction can reach $6dB$~\cite{Complement}. It has been shown that, these limits cannot be surpassed still by {\it successively} utilized the spin-oscillator entanglement and vibrational squeezing~\cite{Affolter2023}.

In this letter, we propose an approach, by simultaneously using the spin-oscillator entanglement and vibrational squeezing, to further surpass the above sensitivity gain over the SQL for the weak electric field measurements. Different from the {\it successive} evolutions used in the previous scheme~\cite{Burd,Affolter2023}, here the initial state preparation (i.e., the entangled and squeezing operations) and the phase accumulation (i.e., the application of the electric field) are implemented {\it simultaneously}, rather than the {\it successively} applied in time. Compared with the previous proposal by using either the spin-oscillator entanglement and squeezing alone~\cite{Gilmore}, the present method can effectively amplify the measured electric field signal and the spin-oscillator entanglement strength, and thus the SQL can be further surpassed for the equal evolution time and mean phonon number. Hopefully, the present scheme with the above integrated advantages can provide a novel approach to implement the desired sub-SQL sensing of electric field.

\begin{figure}[h]
  \centering
  \includegraphics[width=0.45\textwidth]{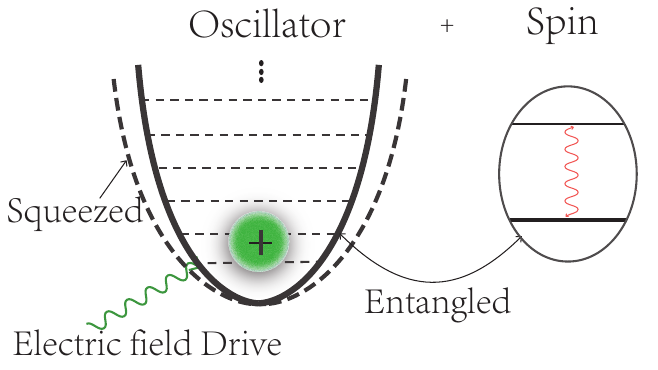}
  \caption{An electric field force driven trapped ion system is used to implement a hybrid interferometer for the sensitive electric field sensing. Here, the system evolves under the simultaneous action of the electric field drive (to be measured), the external vibrational squeezing and the spin-oscillator entanglement.}\label{syt}
\end{figure}

%\section{Model}
{\it Model.---}
Let us consider a hybrid quantum system shown in Fig.~~\ref{syt}, wherein a one-dimensional external linear vibration is coupled to an internal two levels (i.e., a 1/2 spin). The external vibration of the trapped ion is served as the probe to sense the driving electric field force, which can be effectively amplified by squeezing the external vibration. The spin, on the one hand, is used to prepare the spin-oscillator entangled state.
On the other hand, the information of the electric field information encoded in the vibrational state can be transferred to the spin state, which can be detected sensitively with significantly long coherent time~\cite{Julen2020}.

\begin{figure}[h]
  \centering
  \includegraphics[width=0.45\textwidth]{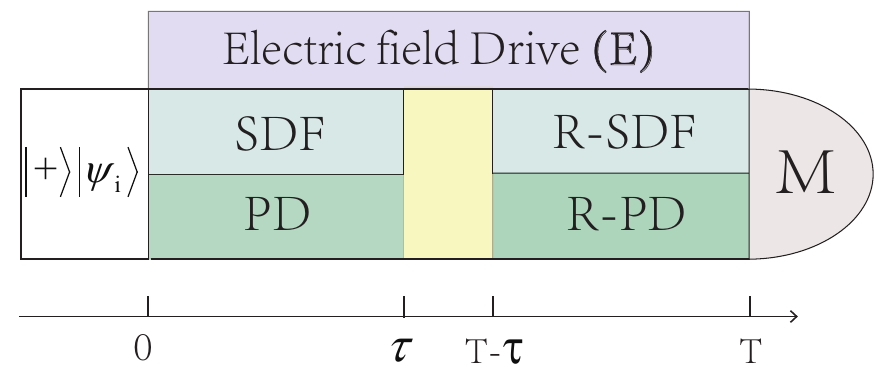}
  \caption{A novel operational protocol of the electric field sensing by simultaneously using spin-oscillator entanglements (generated by the spin-dependent forces (SDFs)) and vibrational squeezings (which are implemented by applying the parametric drives (PDs)).Noted that these operations were successively utilized in the previous schemes.}\label{syt2}
\end{figure}

A procedure designed to implement the desired precise measurement is shown in Fig.~\ref{syt2}, wherein $|+\rangle|\psi_i\rangle$ is the initial state of the system with $|\psi_i\rangle$ being an arbitrary one-dimensional harmonic oscillator state and $|+\rangle=(|\uparrow\rangle+|\downarrow\rangle)/\sqrt{2}$ the superposed internal spin state. After the evolution duration $T$, spin-state projective measurement $"M"$ is performed to implement the sensitive measurement of the applied electric field (i.e., "E"), which displaces the oscillator. Specifically, a spin-dependent force (SDF) and parametric drive (PD), used respectively for the implementations of spin-oscillator entanglement and vibrational squeezing, are simultaneously applied with the electric field to evolve the system within the duration $[0,\tau]$ for the phase accumulation, followed by a free evolution with the duration  $T-2\tau$. Then, the reverse SDF (R-SDF) and PD (R-PD) are applied (with the same duration $\tau$) to disentangle the spin and oscillator while transferring the information of electric field into the spin state, which will be detected later. Noted that the present scheme is basically different from the previous ones~\cite{Affolter2023} wherein the spin-oscillator entanglement, vibrational squeezing operation, and also the detected electric field are applied successively and separated in time.

Physically, the SDF used to deliver the spin-oscillator entanglement is described by the Hamiltonian $\hat{H}_{qe}=(\alpha\hat{a}^\dagger+\alpha^*\hat{a})\sigma_z$,
with $|\alpha|$ being the spin-oscillator coupling strength, and $\sigma_z$ being the Pauli operator acting on the spin. While, the PD used to generate vibrational squeezing is described by the Hamiltonian~\cite{Burd}, $\hat{H}_{qs}=\frac{g}{2}\left(\hat{a}^{\dagger2}e^{i\theta}+\hat{a}^2e^{-i\theta}\right)$,
%\begin{equation}\label{qs}
%  \hat{H}_{qs}=\frac{g}{2}\left(\hat{a}^{\dagger2}e^{i\theta}+\hat{a}^2e^{-i\theta}\right),
%\end{equation}
with $g$ and $\theta$ being the PD strength and phase, respectively. Note that, when $g=0$ (i.e., without squeezing), our model reduces to the case in the Ref~\cite{Gilmore} that use the spin-oscillator entanglement alone; Similarly, when $\alpha=0$ (without the spin-oscillator entanglement), our model reduces to the case treated in Ref.~\cite{Burd} wherein only squeezing on the external oscillator is considered.
The interaction between the detected electric field and the oscillator can be described by the Hamiltonian $\hat{H} _{int}=\eta(\hat{a}^\dagger+\hat{a})$, where $\hat{a}^\dagger(\hat {a})$ is the creation (annihilation) operator acting on the harmonic oscillator, and $\eta=Eqz_0$ denotes the electric filed parameter to be estimated, with $E$,$q$ and $z_0$ being the magnitude of detected electric field, the charge and zero-point fluctuation of ion, respectively.

Specifically, in the interval $(0,\tau)$ the dynamics of the system described by the following Hamiltonian:
\begin{equation}\label{hamilton0}
  \hat{H}_{+}=\eta(\hat{a}+\hat{a}^\dagger )+i\alpha(\hat{a}^\dagger-\hat{a})\sigma_z+\frac{ig}{2}(\hat{a}^{\dagger2}-\hat{a}^2).
\end{equation}
The corresponding evolution operator reads (see Appendix ~\ref{PA}, in detail)
\begin{equation}\label{evo1}
\begin{split}
\hat{U}_{+}(\tau)=&e^{i\phi_0\sigma_z}\hat{D}\left[-\frac{i\eta}{g}\left(e^{-g\tau}-1\right)
+\frac{\alpha\sigma_z}{g}\left(e^{g\tau}-1)\right)\right]\hat{S}(g\tau),
\end{split}
\end{equation}
where $\phi_0=2\eta\alpha\left[\sinh(g\tau)-g\tau\right]/g^2$. With such an evolution, the separable input state $|\psi_s\rangle=|+\rangle|\psi_i\rangle$ can be evolved as a spin-dependent entangled squeezed state:
\begin{equation}\label{psi_eta}
 |\psi(\eta)\rangle=\frac{1}{\sqrt{2}}\left(e^{i\phi_0}|\uparrow\rangle|\psi_+\rangle+e^{-i\phi_0}|\downarrow\rangle|\psi_-\rangle\right),
\end{equation}
with
\begin{equation}
|\psi_{\pm}\rangle=\hat{D}[-\frac{i\eta}{g}\left(e^{-g\tau}-1\right)\pm \frac{\alpha}{g}\left(e^{g\tau}-1)\right)]
\hat{S}(gt)|\psi_s\rangle.
\end{equation}
Noted that the information of the detected electric field has been encoded into the state Eq.~(\ref{psi_eta}), which can be freely evolved within the duration $(\tau, T-\tau)$ described by the evolution operator:  $\hat{U}(\eta)=e^{i\eta(\hat{a}^\dagger+\hat{a})(T-2\tau)}$. Thirdly, the reverse SDF and PD operations are applied simultaneously, with the same duration $\tau$, to disentangle the spin and oscillator while transferring the information of parameter $\eta$ to the spin. This progress can be achieved by the Hamiltonian
\begin{equation}\label{hamilton1}
  \hat{H}_{-}=\eta(\hat{a}+\hat{a}^\dagger )-i\alpha(\hat{a}^\dagger-\hat{a})\sigma_z-\frac{ig}{2}(\hat{a}^{\dagger2}-\hat{a}^2),
\end{equation}
wherein the reverse SDF and the reverse PD are implemented by flipping the spin and adjusting the squeezing phase $\theta=\pi$~\cite{Burd,Gilmore,Affolter2023}, respectively. Correspondingly, the dynamical evolution operator can be expressed as
%\begin{widetext}
\begin{equation}\label{evo2}
\begin{split}
\hat{U}_{-}(\tau)=e^{i\phi_0\sigma_z}\hat{D}\left[\frac{i\eta}{g}\left(e^{g\tau}-1\right)
+\frac{\alpha\sigma_z}{g}\left(e^{-g\tau}-1)\right)\right]\hat{S}^\dagger(g\tau).
\end{split}
\end{equation}
%\end{widetext}
After the above operations with the total duration $T$, the output state of the system reads (see Appendix~(\ref{psi_fa}), in detail)
\begin{equation}\label{psi_f}
\begin{split}
&|\psi_f(\eta)\rangle =\hat{U}_{-}(\tau)\hat{U}(\eta)\hat{U}_+(\tau)|\psi_i\rangle|+\rangle\\
 = &e^{i\phi\sigma_z}\hat{D}\left[\frac{2i\eta}{g}(e^{g\tau}-1)
  +i\eta(T-2\tau)e^{g\tau}\right]|\psi_i\rangle|+\rangle,
\end{split}
\end{equation}
with
\begin{equation}\label{phi_1}
  \phi=4\alpha\eta\left(\frac{e^{g\tau}-1-g\tau}{g^2}+\frac{(e^{g\tau}-1)(T-2\tau)}{2g}\right),
\end{equation}
being the total phase accumulation. Eq.(\ref{psi_f}) indicates that the output state of the system is a spin-oscillator separable quantum states.  For a given $T$, one can properly set the duration $\tau$ to maximize the value of $\phi$. Typically, when $\tau=T/2$, the maximum value $\phi_{es}=4\alpha\eta\left({e^{g\tau}-1-g\tau}\right)/{g^2}$ can be obtained, which provide the optimal phase estimation by performing the spin projective measurement with population:
\begin{equation}\label{population}
  P_{\downarrow}=\frac{1}{2}(1+\cos(\phi_{es})).
\end{equation}
of the spin-state $|\downarrow\rangle$ in the output state~(\ref{psi_f}). The accuracy of such a phase estimation determines the sensitivity of the electric field sensing.

\begin{figure}[htbp]
\centering
\centering
\includegraphics[width=0.48\textwidth]{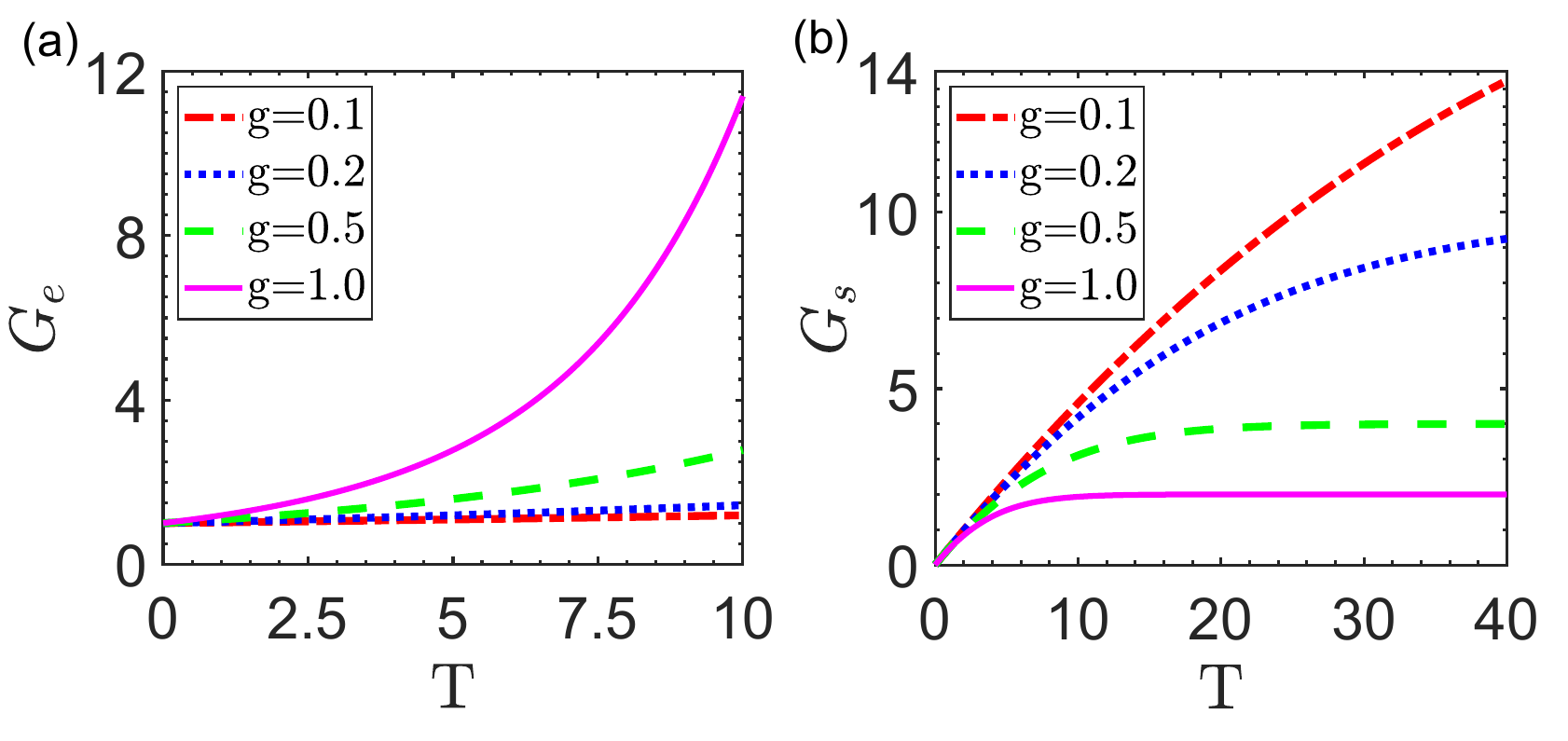}
\caption{ (a) The estimated sensitivity $\Delta\beta=\Delta\eta T$ changes with the mean phonon $\bar{n}$ for different squeezing strength $g$. Here, the SQL (scaling$~1/\sqrt{\bar{n}}$) and HL (scaling$~1/\bar{n}$) are defined in terms of the maximum mean phonon number shown in of Eq.(\ref{averp}). (b) The contour plot of sensitivity $\Delta\beta$ compared with the mean phonon number (blue lines) to optimize the parameter $g$ for the optimal sensitivity. Note that here we have set $\alpha=1$ for convenience, which is the same below without specification.}\label{amplification}
\end{figure}

%\section{Reachable sensitivity of the electric field sensing}
{\it Reachable sensitivity of the electric field sensing.---}
With the spin projective measurement result~(\ref{population}) and the error propagation formula~\cite{Apellaniz2014}, we get the optimal measurement sensitivity
\begin{equation}\label{des}
  \Delta\eta_{es}=\frac{r^2}{\alpha T^2(e^{r}-1-r)},
\end{equation}
of the electric field dependent parameter $\eta$, where $r=gT/2$. Obviously, for $g\rightarrow 0$, i.e., the case without the oscillator squeezing, we have $\Delta\eta_{e}\rightarrow2/\alpha T^2$, which has been demonstrated in Ref.~\cite{Gilmore} using only the spin-oscillator entanglement. While, for $\alpha=0$, i.e., the case that only uses the squeezing, the sensitivity of the estimated $\eta$-parameter reads $\Delta\eta_s={g}/(e^r-1)$~\cite{Burd}, which is related to the squeezing parameter $r$. Interestingly, one can see from Eq.~(\ref{des}) and Fig.~\ref{amplification} that, by simultaneously using the squeezing and entanglement, the sensitivity of the estimated $\eta$-parameter can be amplified by either a factor of $G_s=2\alpha(e^{r}-1-r)/(g(e^{r}-1))$ or the factor of $G_e=2(e^{r}-1-r)/r^2$, compared with that obtained above by only using either the squeezing or entanglement, correspondingly. This implies that one can realize the same sensitivity with the shorter duration by simultaneously using the entanglement and squeezing to speedup the signal accumulation. Specifically, Fig.~\ref{amplification}(b) shows that, the factor $G_e$ can be further amplified by increasing the squeezing strength $g$, even the duration $T$ is fixed. By comparison, for a given $T$, Fig.~\ref{amplification}(a) indicates that the value of the factor $G_s$ is limited by the parameter $2\alpha/g$ and decreases with the increases of $g$. Therefore, for $g\leq 2\alpha$ and the relatively large $T$, the present protocol, i.e., by simultaneously using entanglement and squeezing, possesses the manifest sensitivity advantage over that realized previously by using only entanglement or squeezing.

\begin{figure}[htbp]
\centering
\centering
\includegraphics[width=0.35\textwidth]{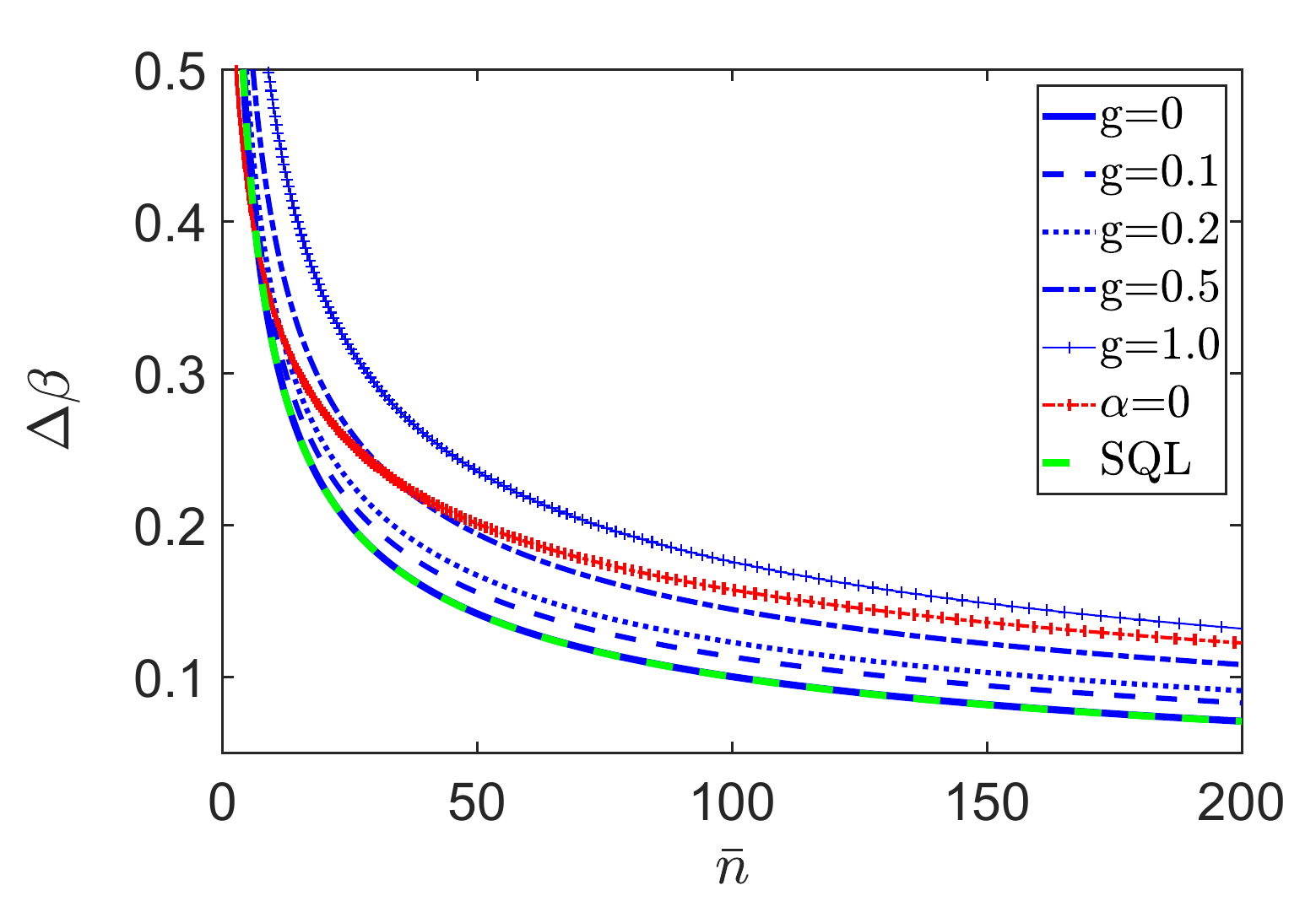}
\caption{ The estimated sensitivity of displacement $\Delta\beta$ changes with the mean phonon $\bar{n}$ for different squeezing strength $g$. Here, the SQL (scaling$~1/\sqrt{\bar{n}}$) is defined in terms of the maximum mean phonon number shown in of Eq.(\ref{averp}). Note that here we have set $\alpha=1$ for convenience, which is the same below without specification.}\label{dtheta_6}
\end{figure}

We now investigate the SQL of the sensitivity for the above electric field sensing, which is related to the mean phonon number of the oscillator~\cite{Caves,Gilmore}. As the mean phonon number is changed during the dynamical evolution, we just consider its maximum value
\begin{equation}\label{averp}
\begin{split}
  \bar{n}_{es}=\eta^2\left( \frac{e^{-r}-1}{g}\right)^2+\alpha ^2\left(\frac{e^{r}-1}{g}\right)^2+\sinh^2(r),
\end{split}
\end{equation}
which is arrived at $T/2$. Note that such a maximum mean phonon number is also available for the case that only uses either entanglement (i.e., $g\rightarrow 0$) or squeezing ($\alpha=0$) Such as, for $g\rightarrow 0$, the above mean phonon number becomes $\bar{n}_e\approx(\alpha T/2)^2$.
By comparing the sensitivity $\Delta\eta_{k}$ $(k=es,e,s)$ with the corresponding mean phonon number $\bar{n}_k$, one can easily verify that $\Delta\eta_{k}>1/(T\sqrt{\bar{n}_{k}})$, which indicates $\Delta\beta_{k}>1/\sqrt{\bar{n}_k}$  with $\Delta\beta_{k}=\Delta\eta_k T$ being the corresponding displacement sensitivity. This implies that, the displacement sensitivity still can not beat the SQL [see Fig.~\ref{dtheta_6}], which is scaling as $1/\sqrt{\bar{n}_k}$ in terms of the mean phonon number $\bar{n}_k$.
%The reason why the present scheme can enhance the sensitivity while can not beat the SQL is that, the simultaneous using of the squeezing the can only amplify the entanglement strength $\alpha$ as $\alpha(e^{r}-1)/g$ (which amplifies the phonon number) meanwhile reduce the electric field parameter $\eta$ as $\eta(e^{-r}-1)/g$.

\begin{figure}[ht]
  \centering
  \includegraphics[width=0.48\textwidth]{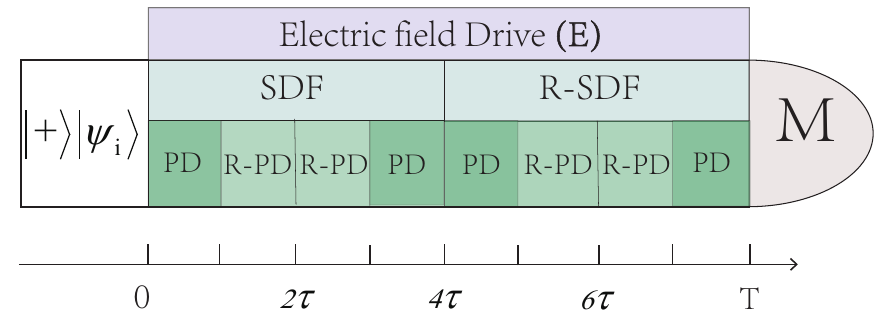}
  \caption{The multi-squeezing sensing protocol. Here multiple squeezing and anti-squeezing operations are applied alternately to simultaneously amplify the electric field parameter $\eta$ and entanglement strength $\alpha$.}\label{msyt}
\end{figure}

{\it Sub-SQL sensing with multiple squeezing protocol (MSP).---}Here we  will show that the sensitivity of the electric field sensing can go beyond the SQL by simultaneously using multiple squeezing as shown in Fig.~\ref{msyt}, wherein the multiple squeezing and anti-squeezing operations are implemented alternately during the same entanglement operation. In this case, the whole procedure can be described as (see Appendix~\ref{PD})
\begin{equation}\label{mse}
\begin{split}
  |\Psi_{f}\rangle &= \hat{V}_{-+}\hat{V}_{--}\hat{V}_{--}\hat{V}_{-+}\hat{V}_{++}\hat{V}_{+-}\hat{V}_{+-}\hat{V}_{++}|\psi_s\rangle\\
  =&\hat{D}\left(\frac{4\sinh(g\tau)}{g}(i\eta+\alpha)\right)\hat{D}\left(\frac{4\sinh(g\tau)}{g}(i\eta-\alpha)\right)|\psi_s\rangle\\
  =& e^{i\Phi\sigma_z}\hat{D}\left(\frac{8i\eta\sinh(g\tau)}{g}\right)|\psi_s\rangle,
  \end{split}
\end{equation}
where $\hat{V}_{\pm,\mp}=e^{i\tau\hat{H}_{\pm,\mp}}$ with
\begin{equation}\label{hamilton_m}
  \hat{H}_{\pm\mp}=\eta(\hat{a}+\hat{a}^\dagger )\pm i\alpha(\hat{a}^\dagger-\hat{a})\sigma_z\mp\frac{ig}{2}(\hat{a}^{\dagger2}-\hat{a}^2),
\end{equation}
and the phase
\begin{equation}\label{mphase}
  \Phi=16\eta\alpha\sigma_z\frac{\sinh^2(g\tau)}{g^2}=\frac{\eta\alpha T^2}{2}\frac{\sinh^2(gT/8)}{(gT/8)^2}\sigma_z.
\end{equation}
From the phase $\Phi$, the electric field parameter $\eta$ can be estimated with the sensitivity
\begin{equation}\label{mdtheta}
  \Delta\eta_m=\frac{2}{\alpha T^2\sinh^2(gT/8)/(gT/8)^2},
\end{equation}
which is amplified by a factor of $\sinh^2(gT/8)/(gT/8)^2$ by the squeezing. In this case, the corresponding maximum mean excited phonon number for this MSP (achieved at evolution time $t=4\tau$, or $t=3\tau$) becomes
\begin{equation}\label{maverp}
  \bar{n}_m\approx \max\left(\left(\frac{\alpha T}{2}\right)^2\frac{\sinh^2(r_m)}{r^2_m},\sinh^2(r_m)\right),
\end{equation}
where $r_m=gT/8$.

\begin{figure}[ht]
  \centering
  \includegraphics[width=0.38\textwidth]{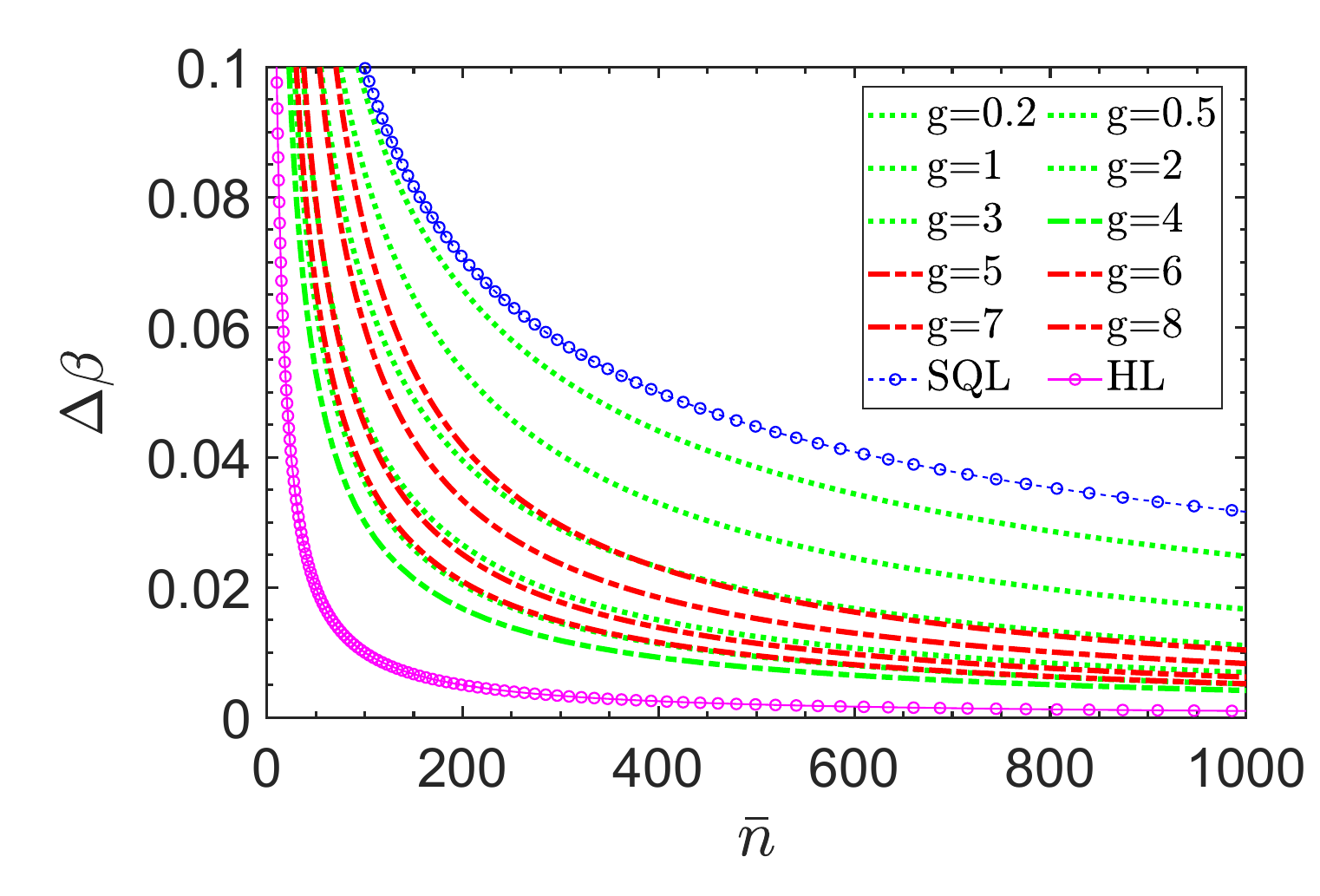}
  \caption{The sensitivity of displacement sensing $\Delta\beta$ and the corresponding SQL (scaling $1/\sqrt{n}$) as well as HL (scaling $1/\bar{n}$) versus the mean phonon number $\bar{n}_m$ for different squeezing parameters $g$. }\label{msf}
\end{figure}

By comparing the sensitivity of Eq.~(\ref{mdtheta}) with the above mean phonon number, one can easily verify that the sensitivity of the electric field sensing as well as the corresponding displacement sensing can go beyond the SQL, i.e., $\Delta\beta<1/\sqrt{\bar{n}}$ is satisfied for arbitrary parameter $g$ as shown in Fig.~\ref{msf}. Specifically, when $g\leq4\alpha$, we have $\bar{n}_m=\left({\alpha T}/{2}\right)^2{\sinh^2(g\tau)}/{(g\tau)^2}$, in which case the sensitivity, $\Delta\beta=r_m/(\sinh(r_m)\sqrt{\bar{n}_m})$, increases with the squeezing strength $g$ for a given $\bar{n}_m$. When $g>4\alpha$, we get $\bar{n}_m\approx \sinh^2(r_m)$, in this case the sensitivity, $\Delta\beta=gr_m/(4\alpha\sinh(r_m)\sqrt{\bar{n}})$,  would decrease rather than increase with $g$ for the given $\bar{n}_m$. Thus, the optimal sensitivity of displacement sensing for the same $\bar{n}_m$ is achieved when $g=4\alpha$, with which the sensitivity is farthest beyond the SQL [see Fig.~\ref{msf}]. Physically, the beaten of the SQL with the present MSP is mainly originated from the simultaneous amplification
of the electric field interaction strength (i.e., parameter $\eta$) and entanglement strength ($\alpha$) of the Hamiltonian by simultaneously utilizing the multiple squeezing operations, which lead to the faster phase accumulation with the same mean phonon number. That is different from the above scheme with the single squeezing operation [see Fig.~\ref{syt2}], wherein only one of the parameters $\eta$ or $\alpha$ can be amplified. Here, the amplification of the parameter $\eta$, which can significantly enhance the sensitivity, has little influence on the increase of the mean phonon number $\bar{n}_m$ since $(\eta\sinh(r_m)/r_m)^2\ll\bar{n}_m$ due to $\eta\ll \alpha$. Note that, the simultaneous amplification of parameters $\alpha$ and $\eta$ can be also achieved by sequentially applying the squeezing and anti-squeezing operations with the Hamiltonian amplification method~\cite{Cosco2021,Arenz2020,Burd2023}, which, however, can not help to further beat the SQL because of the same amplification of the mean phonon number.

\begin{figure}[h]
  \centering
  \includegraphics[width=0.48\textwidth]{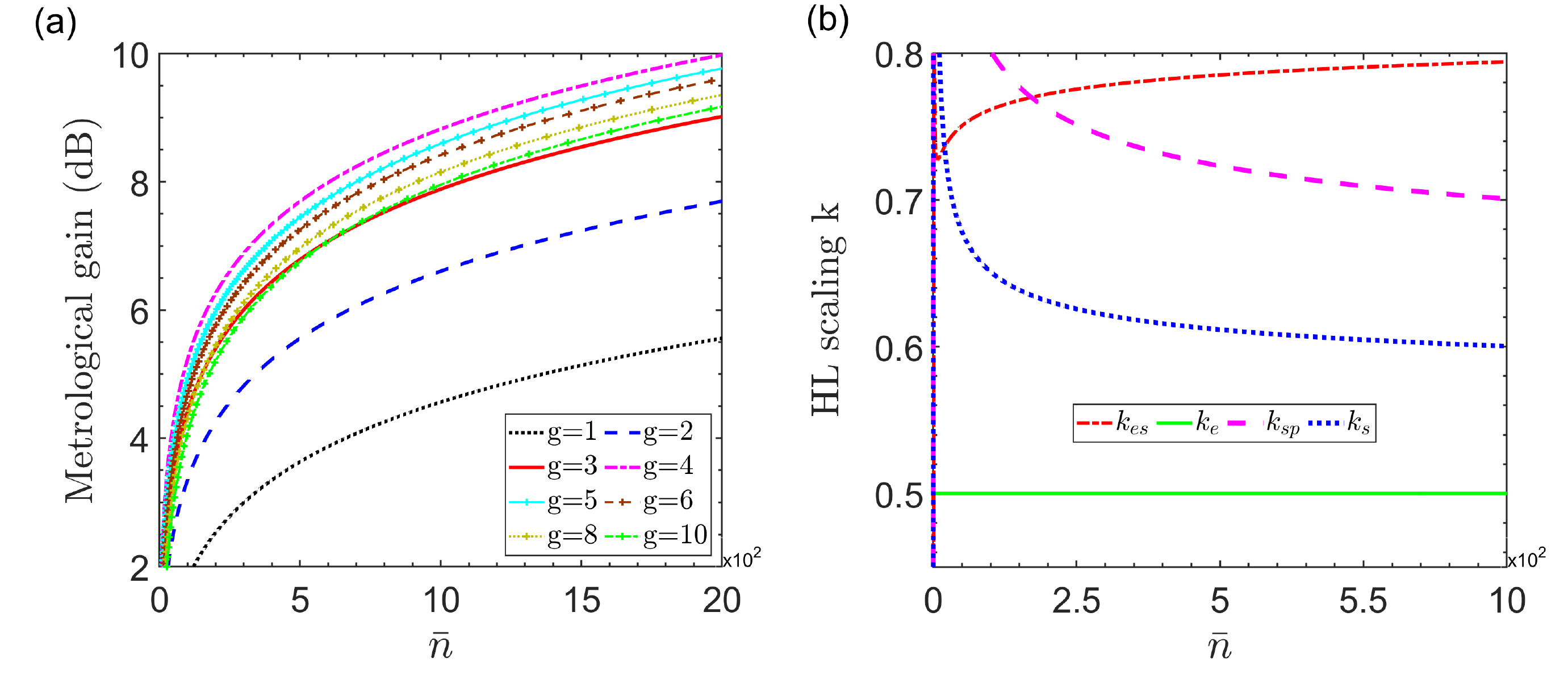}
  \caption{(a) The metrological gains of the sensitivity over the SQL, i.e., $G_{ms}$ and $G_{ml}$ for different $g\leq4\alpha$ and $g>4\alpha$, respectively. (b) The HL scaling, $k_{es}=\ln(\Delta\beta_{es}/\bar{\sqrt{n}_{es}})$, for the present MSP compared with the ones obtained with other schemes, i.e., $k_e=0.5$ of the case that only use entanglement ~\cite{Gilmore}, $k_s$ ($3dB$ gain) of the case that only use squeezing successively~\cite{Burd}, and the upper bound $k_{sp}$ (6dB gain) of the successive evolution that obtained with the QFI. }\label{mgain}
\end{figure}

It is valuable to emphasize that, although a series of the other approaches~\cite{Wolf2019,Burd,Affolter2023} have been demonstrated to beat the SQL by using such as the Fock state, squeezed state and entangled squeezed cat-state as the input states of the hybrid interferometer with trapped ion for the sensitive weak force sensing, the reachable sensitive gain over the relevant SQL can not theoretically surpass $6dB$ for $\bar{n}\gg1$ (the relevant proof is shown in Appendix~\ref{PC}) with much longer evolution duration. Interestingly, here we find that by simultaneously using the entanglement and multiple squeezing operations, any higher metrological gains beyond $6dB$ can be theoretically achieved only if the mean phonon number is sufficiently large. Specifically, with the present MSP, the metrological gains of the sensitivity~(\ref{mdtheta}) over the SQL read
\begin{equation}\label{G_s}
  G_{ms}=10\log\left(\frac{\sinh(gT/8)}{gT/8}\right) (dB).
\end{equation}
and
\begin{equation}\label{G_l}
  G_{ml}=10\log(\frac{4\alpha}{g}\frac{\sinh(gT/8)}{(gT/8)}) (dB),
\end{equation}
for $g\leq4\alpha$ and $g>4\alpha$, respectively. Obviously, both of the above gains $G_{ms}$ and $G_{ml}$ increases with the value $gT$, and can theoretically approach arbitrary value only if $gT$ is sufficiently large. Such as, the gains can surpass $8dB$ when $\bar{n}_m\geq10^3$ and $3\alpha\leq g\leq 10\alpha$ [see Fig.~\ref{mgain} (a)], which indicates the $6dB$ sensitivity with the successive evolution can be overcomed with the present MSP. However, they can not still reach the HL, though much larger gains can be realized with larger phonon excitation,

To investigate how the degree that the sensitivity close to the HL, the HL scaling~\cite{Albarelli2018}, i.e., parameter $k=\log(\Delta\beta)/\log(\bar{n})$, is introduced, which is compared to the ones with previous demonstrated schemes and the optimal one of the successive evolution obtained with the QFI as shown in Fig.~\ref{mgain} (b). One can clearly see that, with the present MSP, the corresponding HL scaling $k_{es}$ is always increasing with $\bar{n}$, thus which is obviously larger than $k_e=0.5$ that uses entanglement only in Ref.~\cite{Gilmore}, $k_s$ that only uses squeezing successively in Ref.~\cite{Burd} (i.e., $3dB$ gain with $k_s\approx0.5+\ln(2)/\ln(\bar{n})$) as well as the theoretical optimized $k_{sp}$ of the successive evolution  (i.e., $6dB$ gain with $k_{sp}\approx0.5+\ln(4)/\ln(\bar{n})$) that obtained by the QFI [see Appendix~\ref{PC}] for relatively large $\bar{n}$.

{\it Conclusions and discussions.---}
In summary, we propose a quantum sensing scheme with a single trapped ion to improve the measurement sensitivity of the applied electric field, by simultaneously utilizing the quantum entanglement and squeezing resources. The basic idea is to simultaneously amplify the electric field strength and the entanglement strength to speedup the phase accumulating. It is shown that, by properly setting the parameters, the sensitivity of the estimated electric field strength can be significantly improved into the sub-SQL regime with higher metrological gain beyond the $6dB$ bound of the previous schemes. Note that, the proposal work also for the SQL defined alternatively by the other forms, typically such as $\propto 1/(2T)$) proposed in Ref.~\cite{Gilmore} in terms of the optimal measurement sensitivity with the coherent state input. In fact, such a definition is independent of the mean phonon number $\bar{n}$. Thus, the SQL can always be beaten by controlling the parameters $\alpha$, $g$, and also the free evolution duration $\tau$. Theoretically, the sensitivity demonstrated here can go far beyond such a SQL, even the HL might be also surpassed for the sufficiently large $g\tau$ or $\alpha$.

It is argued that the experimental realization of the proposal presented here is possible, as the related techniques have been demonstrated with the experimentally-existing trapped ion platform (see, e.g.,~\cite{Leibfried2003,Affolter2023}). For example, the parameters of the entanglement- and squeezing operations and thus the phase accumulation are really controllable by adjusting the driving power and phase. Of course, besides the sensitive sensing of the weak electric field, the present scheme might be also applied to implement the other precise measurements of such as the sensing of rotation~\cite{Stevenson2015,Campbell2017,Feng2022}, as well as the spin-dependent interaction~\cite{Marletto2017,Bose2017,Yant2023} (e.g., the gravitationally induced entanglement) etc.. The robustness of the proposal against various unavoidable noises and the imperfections of the relevant operations, etc. will be discussed elsewhere.

\section*{Acknowledgments}
This work is partially supported by the National Key Research and Development Program of China (NKRDC) under Grant No. 2021YFA0718803 and the National Natural Science Foundation of China (NSFC) under Grant No. 11974290.

%\onecolumn
\newpage
\onecolumngrid
\appendix
%\begin{appendix}
\section{The derivation of evolution operators $\hat{U}_{\pm}$}\label{PA}
Here we give a detail derivation of the evolution operators related to Hamiltonians Eq.(\ref{hamilton0}) and Eq.(\ref{hamilton1}), respectively. Without loss of the generality, let consider the following Hamiltonian
\begin{equation}\label{hamilton2}
\begin{split}
\hat{H}=\beta a^\dagger+\beta^* a+\frac{1}{2}(\xi a^{\dagger2}+\xi^*a^2)
\end{split}
\end{equation}
where $\beta=\eta+i\chi\sigma_z$, $\xi=ge^{i\theta}$. It returns to the Hamiltonian $\hat{H}_+$ shown in Eq.~(\ref{hamilton0}) for $\chi=\alpha$ and $\theta=0$, and the Hamiltonian $\hat{H}_-$ shown in Eq.~(\ref{hamilton1}) for $\chi=-\alpha$ and $\theta=\pi$, respectively. By using the Trotter formula~\cite{Hiai1997}, the evolution operator related to the above Hamiltonian can be expressed as
\begin{equation}
\begin{split}
\hat{U}(t)=& \exp\left(-it(\beta a^\dagger+\beta^* a)-\frac{it}{2}(\xi a^{\dagger2}+\xi^*a^2)\right)\\
=&\lim_{n\rightarrow\infty}\left(\exp\left(\frac{-it}{n}(\beta a^\dagger+\beta^* a)\right)\exp\left(\frac{-it}{2n}(\xi a^{\dagger2}+\xi^*a^2)
\right)\right)^n\\
=&\lim_{n\rightarrow\infty}\Big(\hat{D}(-i\beta t_n)\hat{S}(-i\xi t_n)\Big)^n,
\end{split}
\end{equation}
where $t_n=t/n$, $\hat{D}(x)=\exp(x a^\dagger-x^* a)$, $\hat{S}(x)=\frac{1}{2}(x^*a^2-xa^{\dagger2})$. With
\begin{equation}
\hat{S}^\dagger(x)\hat{a}\hat{S}(x)=\cosh(|x|)\hat{a}-e^{i\arg(x)}\sinh(|x|)\hat{a}^\dagger,~~~~\hat{S}^\dagger(x)\hat{D}(y)\hat{S}(x)=\hat{D}(\cosh(|x|)y+e^{i\arg(x)}\sinh(|x|)y^*),
\end{equation}
where $\gamma(t_n)=-i\beta t_n$ and $\lambda(t_n)=-i\xi t_n$, we have
\begin{equation}
\begin{split}
&\Pi_k \hat{D}(\gamma(t_n))\hat{S}(\lambda(t_n))\cdots\hat{D}(\gamma(t_n))\hat{S}(\lambda(t_n))\hat{D}(\gamma(t_n))\hat{S}(\lambda(t_n))\\
=&\Pi_k \hat{S}(n\lambda(t_n))\left[\hat{S}^\dagger(n\lambda(t_n))\hat{D}(\gamma(t_n))\hat{S}(n\lambda(t_n))\right]
\cdots\left[\hat{S}^\dagger(2\lambda(t_n))\hat{D}
(\gamma(t_n))\hat{S}(2\lambda(t_n))\right]\left[\hat{S}^\dagger(\lambda(t_n))\hat{D}(\gamma(t_n))\hat{S}(\lambda(t_n))\right]\\
&=\Pi_k \hat{S}(n\lambda(t_n))\hat{D}[u_n\gamma(t_n)+v_n\gamma^*(t_n)]\cdots\hat{D}[u_2\gamma(t_n)+v_2\gamma^*(t_n)]\hat{D}[u_1\gamma(t_n)+v_1\gamma^*(t_n)],
\end{split}
\end{equation}
where $u_k=\cosh(|k\xi t_n|)$,$v_k=e^{i\theta}\sinh(|k\xi t_n|)$,$u=\cosh(|\xi t|)$, $v=e^{i\theta}\sinh(|\xi t|)$,and $\theta=\arg(-i\xi)$. As a consequence, we get
\begin{equation}
\begin{split}
\hat{U}(t)=&\lim_{n\rightarrow\infty}\left(\exp\left(\frac{-it_n}{n}(\beta a+\beta^*a^\dagger)\right)\exp\left(\frac{-it_n}{2n}(\xi a^{\dagger2}+\xi^*a^2)
\right)\right)^n\\
=&\lim_{n\rightarrow\infty}\hat{S}(n\lambda(t_n))\Pi_{k=1}^n\hat{D}(u_k\gamma(t_n)+v_k\gamma^*(t_n)).
\end{split}
\end{equation}
Furthermore, by using the relationship $\hat{D}(x)\hat{D}(y)=e^{i\Im(xy^*)}\hat{D}(x+y)$, we can also obtain
\begin{equation}
\lim_{n\rightarrow\infty}\Pi_{k=1}^nD(u_k\gamma(t_n)+v_k\gamma^*(t_n))=
\exp(i\phi)D(\gamma(t_n)\sum_{k=1}^{\infty}u_k+\gamma^*(t_n)\sum_{k=1}^\infty v_k),
\end{equation}
where $\phi=\lim_{n\rightarrow\infty}\phi_n$ with
\begin{equation}
i\phi_n=\frac{1}{2}\sum_{k=2}^n\left[\Big(u_k\gamma(t_n)+v_k\gamma^*(t_n)\Big)\sum_{j=1}^{k-1}
\Big(u_j\gamma^*(t_n)+v_j^*\gamma(t_n)\Big)\right]-h.c.
\end{equation}
As
\begin{equation}
\begin{split}
\sum_{j=1}^{k-1}u_j=&\sum_j\frac{1}{2}(e^{jgt_n}+e^{-jgt_n})
=\frac{1}{2}(\frac{e^{kgt_n}-1}{e^{gt_n}-1}
+\frac{e^{-kgt_n}-1}{e^{-gt_n}-1})
\approx\frac{1}{gt_n}\sinh(kgt_n),
\end{split}
\end{equation}
and
\begin{equation}
\begin{split}
\sum_{j=1}^{k-1}v_j=\sum_j\frac{e^{i\theta_1}}{2}(e^{jgt_n}-e^{-jgt_n})
=\frac{e^{i\theta_1}}{2}(\frac{e^{kgt_n}-1}{e^{gt_n}-1}
-\frac{e^{-kgt_n}-1}{e^{-gt_n}-1})
\approx\frac{e^{i\theta_1}}{gt_n}(\cosh(kgt_n)-1),
\end{split}
\end{equation}
where $\theta_1=\theta-\pi/2$, one can further get
\begin{equation}\label{phi}
\begin{split}
i\phi&=\sum_{k=2}^\infty\frac{1}{2gt_n}\Big[\Big(\cosh(kgt_n)\gamma(t_n)+e^{i\theta_1}\sinh(kgt_n)\gamma^*(t_n)\Big)
\Big(\sinh(kgt_n)\gamma^*(t_n)+e^{-i\theta_1}(\cosh(kgt_n)-1)\gamma(t_n)\Big)\Big]-h.c\\
&=\sum_{k=2}^\infty\frac{-1}{2gt_n}\Big[\sinh^2(kgt_n)\gamma^{*2}(t_n)e^{i\theta_1}+
\cosh(kgt_n)[\cosh(kgt_n)-1]\gamma^2(t_n)e^{-i\theta_1}\Big]-h.c\\
&=\sum_{k=2}^\infty\frac{-1}{2gt_n}\Big[i(2\eta\chi\cos(\theta_1)\sigma_z+(\eta^2-\chi^2)\sin(\theta_1))(\cosh(gt)-1)t_n^2\Big]-h.c\\
&=\frac{-i}{2g}\int_0^{T/2}dt\Big[\big(2\eta\chi\cos(\theta_1)\sigma_z+(\eta^2-\chi^2)\sin(\theta_1)\big)(\cosh(gt)-1)\Big]-h.c\\
&=i\frac{2\eta\chi\cos(\theta_1)\sigma_z+(\eta^2-\chi^2)\sin(\theta_1)}{g^2}\Big[
\sinh(\frac{gT}{2})-\frac{gT}{2}\Big].
\end{split}
\end{equation}
Therefore, for $\chi=\alpha$ and $\theta=\pi/2$, i.e.,$\xi=ig$, the evolution operator $\hat{U}(t)$ shown in Eq.~(A5) reduces to
\begin{equation}\label{u2}
\begin{split}
\hat{U}_+(t)=&\lim_{n\rightarrow\infty}\hat{S}(n\lambda(t_n))\hat{D}(\sum_{k=1}^nu_{k}\gamma(t_n)-\sum_{k=1}^nv_k\gamma^*(t_n))\\
=&\exp(i\phi_0)\hat{S}({gT}/{2})\hat{D}\Big[\frac{i}{g}\left(\sinh(gT/2)\beta
+(\cosh(gT/2)-1)\beta^*\right)\Big]\\
=&\exp(i\phi_0)\hat{S}({gT}/{2})\hat{D}\left[\frac{i\eta}{g}\left(\sinh(gT/2)
+\cosh(gT/2)-1)\right)
+\frac{\alpha\sigma_z}{g}\left(\sinh(gT/2)
-\cosh(gT/2)-1)\right))\right]\\
=&\exp(i\phi_0)\hat{S}({gT}/{2})\hat{D}\left[\frac{i\eta}{g}(e^{gT/2}-1)
-\frac{\alpha\sigma_z}{g}(e^{-gT/2}-1)\right]\\
=&\exp(i\phi_0)\hat{D}\left[\frac{-i\eta}{g}(e^{-gT/2}-1)
-\frac{\alpha\sigma_z}{g}(e^{gT/2}-1)\right]\hat{S}({gT}/{2}),
\end{split}
\end{equation}
 which is the just result of Eq.~(\ref{evo1}), with $\phi=2\alpha\eta(\sinh(gT/2)-gT/2)\sigma_z$.
Similarly, for $\chi=-\alpha$ and $\theta=-\pi/2$, the evolution operator $\hat{U}(t)$ shown in Eq.~(A5) reduces to
\begin{equation}\label{u1}
\begin{split}
\hat{U}_-(t)=&\lim_{n\rightarrow\infty}\hat{S}^\dagger(n\lambda(t_n))\hat{D}(\sum_{k=1}^nu_{k}\gamma(t_n)+\sum_{k=1}^nv_k\gamma^*(t_n))\\
=&\exp(i\phi_0)\hat{S}^\dagger({gT}/{2})\hat{D}\Big[\frac{i}{g}\left(\sinh(gT/2)\beta
-(\cosh(gT/2)-1)\beta^*\right)\Big]\\
=&\exp(i\phi_0)\hat{S}^\dagger({gT}/{2})\hat{D}\left[\frac{i\eta}{g}\left(\sinh(gT/2)
-\cosh(gT/2)+1)\right)
+\frac{\alpha\sigma_z}{g}\left(\sinh(gT/2)
+\cosh(gT/2)-1)\right))\right]\\
&=\exp(i\phi_0)\hat{S}^\dagger({gT}/{2})\hat{D}\left[\frac{-i\eta}{g}(e^{-gT/2}-1)
+\frac{\alpha\sigma_z}{g}(e^{gT/2}-1)\right]\\
&=\exp(i\phi_0)\hat{D}\left[\frac{i\eta}{g}(e^{gT/2}-1)
-\frac{\alpha\sigma_z}{g}(e^{-gT/2}-1)\right]\hat{S}^\dagger({gT}/{2}),
\end{split}
\end{equation}
which is the just result of Eq.~(\ref{evo2}).
By combining the Eq.~(\ref{u1}) and Eq~(\ref{u2}), the final evolution state, i.e., Eq.~(\ref{psi_f}), can be obtained as
\begin{equation}\label{psi_fa}
  \begin{split}
    |\psi_f(\eta)\rangle =&\hat{U}_{-}(\tau)\hat{U}(\eta)\hat{U}_+(\tau)|\psi_s\rangle\\
  = & e^{2i\phi_0}\hat{D}\left[\frac{i\eta}{g}(e^{gT/2}-1)
-\frac{\alpha\sigma_z}{g}(e^{-gT/2}-1)\right]\hat{S}^\dagger({gT}/{2})\hat{U}_\eta
\hat{S}({gT}/{2})
 \hat{D}\left[\frac{i\eta}{g}(e^{gT/2}-1)
+\frac{\alpha\sigma_z}{g}(e^{-gT/2}-1)\right]|\psi_s\rangle\\
  = &  e^{2i\phi_0}\hat{D}\left[\frac{i\eta}{g}(e^{-gT/2}-1)
-\frac{\alpha\sigma_z}{g}(e^{-gT/2}-1)\right]\hat{D}(i\eta e^{gT/2}(T-2\tau))
\hat{D}\left[\frac{i\eta}{g}(e^{gT/2}-1)
+\frac{\alpha\sigma_z}{g}(e^{-gT/2}-1)\right]|\psi_s\rangle\\
=& e^{i\phi_1}\hat{D}\left[\frac{i\eta}{g}(e^{gT/2}-1)\right]\hat{D}\left[\frac{-\alpha\sigma_z}{g}(e^{gT/2}-1)\right]\hat{D}(i\eta e^{gT/2}(T-2\tau))
 \hat{D}\left[\frac{\alpha\sigma_z}{g}(e^{-gT/2}-1)\right]\hat{D}\left[\frac{i\eta}{g}(e^{gT/2}-1)\right]|\psi_s\rangle\\
  = & e^{i\phi}\hat{D}\left[\frac{2i\eta}{g}(e^{g\tau}-1)+i\eta(T-2\tau)e^{g\tau}\right]|\psi_s\rangle,
\end{split}
\end{equation}
where $\phi_1=2\phi_0+2\alpha T^2\sigma_z\sinh(gT/4)^2/(gT/2)^2$ and $\phi=4\alpha\eta\left(\frac{e^{g\tau}-1-g\tau}{g^2}+\frac{(e^{g\tau}-1)(T-2\tau)}{2g}\right)\sigma_z$.

\section{The optimal sensitivity with the successive evolution}\label{PC}
As the optimal sensitivity of an arbitrary measurement with quantum state $|\psi\rangle$ is bounded by the Cram$\acute{e}$r-Rao inequality~\cite{Braunstein1994}
\begin{equation}
\Delta(\eta)\leq \frac{1}{\sqrt{F_Q(|\psi\rangle)}},
\end{equation}
with
\begin{equation}\label{QFI}
  F_Q(|\psi\rangle)=4\left[\langle\psi|H_0^2|\psi\rangle-\langle\psi|H_0|\psi\rangle^2\right],
\end{equation}
being the quantum Fisher information (QFI) of the state $|\psi\rangle$. Here, $H_0$ is the generator and reads $H_0=i(\hat{a}^\dagger-\hat{a})$ here for the electric field sensing.
Firstly, for the Fock state $|\psi_F\rangle=|N\rangle$ input, we have
can be calculated as
\begin{equation}\label{Fock}
\begin{split}
F_Q(|N\rangle)=&4(\langle N|-(\hat{a}^\dagger-\hat{a})^2|N\rangle-\langle N|(\hat{a}^\dagger-\hat{a})|N\rangle^2)\\
=& -4\langle N|(\hat{a}^\dagger-\hat{a})^2|N\rangle = 8N+4.
\end{split}
\end{equation}
Therefore, the optimal measurement sensitivity for the Fock state input is $1/\sqrt{F_Q(|N\rangle)}=1/(2\sqrt{2N})$. Similarly, for the squeezed state $|\xi\rangle=\hat{S}(\xi)|0\rangle$ input, the corresponding QFI can be obtained as
\begin{equation}\label{Sq}
\begin{split}
F_Q(|\xi\rangle)=&4(\langle r|-(\hat{a}^\dagger-\hat{a})^2|r\rangle-\langle r|(\hat{a}^\dagger-\hat{a})|r\rangle^2)\\
=& -4\langle r|(\hat{a}^\dagger-\hat{a})^2|r\rangle = 4|u+v|^2=4e^{2r},
\end{split}
\end{equation}
where $u=\cosh(r)$, $v=\sinh(r)$. Obviously, the optimal sensitivity with the squeezed state input is $e^{-r}/2\approx1/4\sqrt{\bar{n}}$ (with $\bar{n}=\sinh^2(r)$ being the mean phonon number), which is about $6dB$ enhancement over the SQL. However, only half of this optimal sensitivity, i.e., $3dB$ enhancement (with sensitivity $e^{-r}$), has been experimentally realized~\cite{Burd} by using the reverse squeezing technique. This implies that, how to realize the theoretically optimal sensitivity $e^{-r}/2$ is still a challenge for the practical experiments.

Thirdly, for an arbitrary input state of the form $|\psi_1\rangle=(|\downarrow\rangle|\psi_+\rangle+|\uparrow\rangle|\psi_-\rangle)/\sqrt{2}$, the corresponding QFI can be also calculated as~\cite{Feng2023},
\begin{equation}\label{Beq4}
    \begin{split}
    F_Q(|\psi_1\rangle)&=2\Big[(\Delta\hat{H} _{0|\psi_+\rangle})^2+(\Delta\hat{H}_{0|\psi_-\rangle})^2\Big]+\left(\langle\psi_+|\hat{H}_0|\psi_+\rangle-\langle\psi_-|\hat{H}_0|\psi_-\rangle\right)^2\\
    &=2\left(\langle\psi_+|\hat{H}_0^2|\psi_+\rangle+\langle\psi_-|\hat{H}_0^2|\psi_-\rangle\right)-\left(\langle\psi_+|\hat{H}_0|\psi_+\rangle-\langle\psi_-|\hat{H}_0|\psi_-\rangle\right)^2\\
    &\leq \max\left(4\langle\psi_+|\hat{H}_0^2|\psi_+\rangle,4\langle\psi_-|\hat{H}_0^2|\psi_-\rangle\right),
  \end{split}
\end{equation}
with $(\Delta\hat{H}_{0|\psi_j\rangle})^2=\langle\psi_j|\hat{H}_0^2|\psi_j\rangle-\langle\psi_j|\hat{H}_0|\psi_j\rangle^2$, $j=+,-$. For $H_{0}=i(\hat{a}^\dagger-\hat{a})$, we have
\begin{equation}
  \begin{split}
  \langle\psi_+|\hat{H}_0^2|\psi_+\rangle=-\langle\psi_+|\hat{a}^{\dagger2}+\hat{a}^2
  -\hat{a}^\dagger\hat{a}-\hat{a}\hat{a}^\dagger|\psi_+\rangle\geq0,
  \end{split}
\end{equation}
which indicates that $\langle\hat{a}^{\dagger2}+\hat{a}^2\rangle\leq \langle\hat{a}^\dagger\hat{a}+\hat{a}\hat{a}^\dagger\rangle$. Thus,
\begin{equation}\label{Beq6}
  \langle\psi_+|\hat{H}_0^2|\psi_+\rangle\leq\langle 2\hat{a}^\dagger\hat{a}+2\hat{a}\hat{a}^\dagger\rangle=4\langle
  \hat{a}^\dagger\hat{a}\rangle+2.
\end{equation}
By substituting Eq.~(\ref{Beq6}) into Eq.~(\ref{Beq4}), we can get
\begin{equation}\label{Beq7}
  F_Q(|\psi_1\rangle)\leq16\bar{n}+8,
\end{equation}
which implies that the optimal sensitivity is bounded by ${1}/{F_Q(|\psi_1\rangle)}={1}/{4\sqrt{\bar{n}+1/2}}$. It corresponds to the gain
\begin{equation}
  G_{s} = 10\log\left(4+\frac{2\sqrt{2}}{\sqrt{\bar{n}}}\right),
\end{equation}
and thus about $6dB$ enhancement over the SQL, for $\bar{n}\gg1$. This is the proof of the argument in Ref.~[22]. It is value to emphasize that, the result shown in Eq.~(\ref{Beq7}) can be applicable to all the input states, typically including such as the Fock state, squeezed state and even the spin-dependent entangled states, etc..
\section{The derivation of the evolution operator of MSP}\label{PD}
In the section a derivation of the final state of Eq.~(\ref{mse}) is given. Firstly, we can directly obtained $\hat{U}_{+-}(\tau)=\hat{U}_{+}(\tau/2)$  and $\hat{V}_{-+}(\tau)=\hat{U}_{-}(\tau/2)$ with $\hat{U}_{+}$ being given in Eq.~(\ref{u1}) and Eq.~\ref{u2}, respectively. Also, according the relationship of Eq.~\ref{u1}, by replacing $g$ with $-g$ and $T/2$ with $\tau$ in the Eq.~(\ref{u1}), the evolution operator of $\hat{V}_{++}(\tau)$ is obtained as
\begin{equation}
  \hat{V}_{++}(\tau)=\exp(i\phi_m)\hat{D}\left[\frac{-i\eta}{g}(e^{-g\tau}-1)
+\frac{\alpha\sigma_z}{g}(e^{g\tau}-1)\right]\hat{S}^\dagger(g\tau).
\end{equation}
Above the phase $\phi_m$ is obtained according to the Eq.~(\ref{phi}) as
\begin{equation}
  \phi_m = \frac{4\alpha\eta\sigma_z}{g^2}\left(\sinh(g\tau)-g\tau\right),
\end{equation}
which is cancelled by the phase $-\phi_m$ of the evolution $\hat{V}_{+-}$. Similarly, by replacing $\alpha$ with $-\alpha$ and replacing $T/2$ with $\tau$ of the Eq.~(\ref{u1}), the evolution operator of $\hat{V}_{--}$ is obtained as
\begin{equation}
  \hat{V}_{--}(\tau)=\exp(-i\phi_m)\hat{D}\left[\frac{-i\eta}{g}(e^{-g\tau}-1)
-\frac{\alpha\sigma_z}{g}(e^{g\tau}-1)\right]\hat{S}(g\tau).
\end{equation}
Then the total evolution operator in the duration (0,T/2) is obtained as
\begin{equation}\label{evo12}
\begin{split}
 \hat{V}_{++}(\tau)\hat{V}_{+-}(\tau)&=e^{i\phi_m}e^{-i\phi_m}\hat{D}\left[\frac{-i\eta}{g}(e^{-g\tau}-1)
+\frac{\alpha\sigma_z}{g}(e^{g\tau}-1)\right]\hat{S}^\dagger(g\tau)\hat{D}\left[\frac{i\eta}{g}(e^{g\tau}-1)
-\frac{\alpha\sigma_z}{g}(e^{-g\tau}-1)\right]\hat{S}(g\tau)\\
&=\hat{D}\left[\frac{-i\eta}{g}(e^{-g\tau}-1)
+\frac{\alpha\sigma_z}{g}(e^{-g\tau}-1)\right]\hat{D}\left[\frac{-i\eta}{g}(e^{-g\tau}-1)
+\frac{\alpha\sigma_z}{g}(e^{g\tau}-1)\right]\\
&=\hat{D}\left[\frac{-2i\eta}{g}(e^{-g\tau}-1)
+\frac{2\alpha\sigma_z}{g}(e^{g\tau}-1)\right].
\end{split}
\end{equation}
Similarly, we have
\begin{equation}\label{evo34}
\begin{split}
 \hat{V}_{+-}(\tau)\hat{V}_{++}(\tau)&=e^{-i\phi_m}e^{i\phi_m}\hat{D}\left[\frac{i\eta}{g}(e^{g\tau}-1)
-\frac{\alpha\sigma_z}{g}(e^{-g\tau}-1)\right]\hat{S}(g\tau)\hat{D}\left[\frac{-i\eta}{g}(e^{-g\tau}-1)
+\frac{\alpha\sigma_z}{g}(e^{g\tau}-1)\right]\hat{S}^\dagger(g\tau)\\
&=\hat{D}\left[\frac{i\eta}{g}(e^{g\tau}-1)
-\frac{\alpha\sigma_z}{g}(e^{-g\tau}-1)\right]\hat{D}\left[\frac{i\eta}{g}(e^{g\tau}-1)
-\frac{\alpha\sigma_z}{g}(e^{-g\tau}-1)\right]\\
&=\hat{D}\left[\frac{2i\eta}{g}(e^{g\tau}-1)
-\frac{2\alpha\sigma_z}{g}(e^{-g\tau}-1)\right].
\end{split}
\end{equation}
By combining the Eq.~(\ref{evo12}) and Eq.~(\ref{evo34}), we get the evolution operator for duration $(0,T/2)$
\begin{equation}\label{evo_a}
\begin{split}
\hat{V}_{+-}(\tau)\hat{V}_{++}(\tau)\hat{V}_{++}(\tau)\hat{V}_{+-}(\tau)&=\hat{D}\left[\frac{-2i\eta}{g}(e^{-g\tau}-1)
+\frac{2\alpha\sigma_z}{g}(e^{g\tau}-1)\right]\hat{D}\left[\frac{2i\eta}{g}(e^{g\tau}-1)
-\frac{2\alpha\sigma_z}{g}(e^{-g\tau}-1)\right]\\
&=e^{i\phi_{m1}}\hat{D}\left[\frac{4i\eta\sinh(g\tau)}{g}
+\frac{4\alpha\sigma_z\sinh(g\tau)}{g}\right],
\end{split}
\end{equation}
where $\phi_{m1}=-4\alpha\eta\sigma_z(e^{r}-1)^2(1-e^{-2r})$. By replacing $\alpha$ of the above Eq.~(\ref{evo_a}) with $-\alpha$, we get the corresponding evolution operator for the duration $(T/2,T)$
\begin{equation}\label{evo_b}
\begin{split}
\hat{V}_{--}(\tau)\hat{V}_{-+}(\tau)\hat{V}_{-+}(\tau)\hat{V}_{--}(\tau)&=\hat{D}\left[\frac{-2i\eta}{g}(e^{-g\tau}-1)
-\frac{2\alpha\sigma_z}{g}(e^{g\tau}-1)\right]\hat{D}\left[\frac{2i\eta}{g}(e^{g\tau}-1)
+\frac{2\alpha\sigma_z}{g}(e^{-g\tau}-1)\right]\\
&=e^{-i\phi_{m1}}\hat{D}\left[\frac{4i\eta\sinh(g\tau)}{g}
-\frac{4\alpha\sigma_z\sinh(g\tau)}{g}\right].
\end{split}
\end{equation}
Note that the above $-\phi_{m1}$ is just cancelled with $\phi_{m1}$ of Eq.~(\ref{evo_a}).
Therefore, the total evolution operator for the whole duration $(0,T)$ is
\begin{equation}\label{evo_t}
\begin{split}
&\hat{V}_{--}(\tau)\hat{V}_{-+}(2\tau)\hat{V}_{--}(\tau)\hat{V}_{+-}
(\tau)\hat{V}_{++}(2\tau)\hat{V}_{+-}(\tau)\\
&=e^{-i\phi_{m1}}e^{i\phi_{m1}}\hat{D}\left[\frac{4i\eta\sinh(g\tau)}{g}
-\frac{4\alpha\sigma_z\sinh(g\tau)}{g}\right]\hat{D}\left[\frac{4i\eta\sinh(g\tau)}{g}
+\frac{4\alpha\sigma_z\sinh(g\tau)}{g}\right]\\
&=e^{i\Phi\sigma_z}\hat{D}\left(\frac{8i\eta\sinh(g\tau)}{g}\right),
\end{split}
\end{equation}
where $\Phi=\frac{\alpha T^2}{2}\frac{\sinh^2(gT/8)}{(gT/8)^2}$. With the above total evolution operator, the final evolved state after the whole dynamical evolution for duration $(0,T)$ can be obtained as
\begin{equation}\label{Amse}
\begin{split}
  |\Psi_{f}\rangle &= \hat{V}_{-+}\hat{V}_{--}\hat{V}_{--}\hat{V}_{-+}\hat{V}_{++}\hat{V}_{+-}\hat{V}_{+-}\hat{V}_{++}|\psi_s\rangle\\
  =&\hat{D}\left(\frac{4\sinh(g\tau)}{g}(i\eta+\alpha)\right)\hat{D}\left(\frac{4\sinh(g\tau)}{g}(i\eta-\alpha)\right)|\psi_s\rangle\\
  =& e^{i\Phi\sigma_z}\hat{D}\left(\frac{8i\eta\sinh(g\tau)}{g}\right)|\psi_s\rangle,
  \end{split}
\end{equation}
which is just the result of the Eq.~(\ref{mse}).

\begin{thebibliography}{99}
\bibitem{Complement} Here the SQL is defined in terms of the maximum mean phonon number excited during the whole evolution procedure. The experimental demonstation of the displacement measurement sensitivity is $\Delta\beta=e^{-r}\approx{1/2\sqrt{\bar{n}}}$, where $\bar{n}=\sinh^2(r)$. It is a $3dB$ gain over the SQL. In fact, for the squeezed state $|\xi\rangle$ input, the theoretically reachable
sensitivity gain is $6dB$ gain. We provide such a proof in the Appendix C.
\end{thebibliography}
\end{document}